\documentstyle[12pt,qqa4lart]{article}
\input tcilatex

\QQQ{Language}{
American English
}

\begin{document}

\author{F.\ Lalo\"e  \\ %EndAName
Laboratoire Kastler Brossel \thanks{%
UA associ\'ee au CNRS no 18, laboratoire associ\'e \`a l'Universit\'e Pierre
et Marie Curie.}}
\title{Ursell operators in statistical physics I:\\Generalizing the Beth
Uhlenbeck
formula. }
\maketitle

\begin{abstract}
The Beth Uhlenbeck formula gives an exact (quantum) expression of the second
virial correction to the equation of state of a dilute gas.\ We show how
this result can be extended to arbitrary degeneracy. For this purpose we
develop a formalism based on the use of Ursell operators, which contain no
symmetrization in themselves (they can be used for both a system of
identical particles or an auxiliary system of distinguishable particles). A
concise expression generalizing the Beth Uhlenbeck formula is obtained,
which is equally valid for bosons and fermions. The formalism is not
essentially limited to dilute gases, but can also be applied to the study of
denser systems, for instance those undergoing a Bose Einstein transition;
this will be the subject of a forthcoming article.
\end{abstract}

\section{Introduction}

The Beth Uhlenbeck formula\cite{BU}\cite{Huang-BU} gives an exact expression
of the second virial coefficient as a function of all collision phase shifts
associated with the interaction potential.\ This is a remarkable result
which, in a concise formula including no phenomenological constant, relates
macroscopic and microscopic quantities: on one hand the density, the
temperature, and the pressure of a gas, on the other the quantum phase
shifts, which are exactly calculable from the Schr\"odinger equation. A
closer inspection shows that the pressure correction is the sum of two
contributions, one of pure statistics, and a second arising from the
interactions; the former is actually nothing but the first order term of the
expansion in $n\lambda ^3$ of the ideal gas equation of state.\ It is
therefore not surprising that the validity of the formula should require two
conditions, that the gas be dilute with respect to both degeneracy effects ($%
n\lambda ^3\ll 1$) and interaction effects as well ($nb^3\ll 1$, where $b$
is the range of the interaction potential). A natural question is whether
one can release the first of these conditions and study gaseous systems over
a larger range, going continuously from the classical region where $n\lambda
^3\ll 1$ to the quantum region where this quantity is comparable to one.\
One would just assume that the gas remains always dilute in terms of the
interactions. As a matter of fact, most classical textbooks on quantum
statistical mechanics include a study of gaseous systems that are degenerate
and imperfect.\ The methods they use are based on various forms of
perturbation theory; for instance they introduce an auxiliary
pseudopotential which replaces the real interaction potential\cite
{Huang-dilues}, or they make a direct substitution of a collision $T$ matrix
for this potential \cite{Landau-dilues}.\ This well known approaches are,
nevertheless, different from that of Beth and Uhlenbeck whose calculation is
exact (of course they limit the calculation to the first density correction
to the equation of state, but this is precisely the definition of the second
virial correction). In essence, the cluster expansion method they use is a
method providing directly density expansions, as opposed to interaction
expansions, which require at least partial resummations of diagrams in
powers of the interaction potential \cite{Galitskii} to reconstruct the $T$
matrix from $V$ and to obtain density expansions. Actually, the exact
treatment of the relative motion of two particles is explicit in the Beth
Uhlenbeck formula, which contains a thermal exponential of the two particle
hamiltonian ensuring that thermal equilibrium has been reached, in other
words that the binary correlations between the particles are treated
properly.\ Technically, a characteristic of this approach is that no
expression which becomes infinite for repulsive hard cores is ever written,
even at some intermediate stage. It therefore seems to be an attractive
possibility to try and extend the method to degenerate systems, if only to
compare the result of both methods, interaction and density expansions. This
is the subject of the present article; we show that concise and exact
formulas can indeed be written that provide a natural generalization of the
Beth Uhlenbeck formula, without requiring complicated algebra.

For this purpose, we will make use of a generalization of the quantum Ursell%
\footnote{%
The initial introduction of the Ursell functions was made by him within
classical statistical mechanics\cite{Ursell}, and generalized ten years
later to quantum mechanics by Kahn and Uhlenbeck.} cluster functions
introduced by Kahn and Uhlenbeck \cite{Kahn-Uhlenbeck}: Ursell operators
which intrinsically do not contain statistics (as opposed to the usual
quantum Ursell functions \cite{Huang-chap14}\cite{Pathria}, which are
already symmetrized).\ The action of the operators is defined, not only in
the space of symmetric or antisymmetric states of the real system, but also
in the larger space of an auxiliary system of distinguishable particles. If
the system is dilute in terms of the interactions but not of statistics,
this makes it possible to limit the calculation to low order Ursell
operators while, with the usual (fully symmetrized) Ursell functions, one
would need to include higher and higher orders with increasing degeneracy%
\footnote{%
As discussed in \cite{FL-1}, when a Bose Einstein condensation takes place,
the Ursell operators have no singular variation, which is another way to see
that they are not sensitive to degeneracy by themselves.}.\ Of course, this
method implies that we have to give up well-known advantages of the
formalism of second quantization, but this is the price to pay for the
treatment of interactions and statistics in completely independent steps. It
also means that an explicit symmetrization of the wave functions becomes
indispensable at some point and, moreover, that no approximation whatsoever
can be made at this step: otherwise the possibility of treating strongly
degenerate systems would be lost.\ Fortunately it turns out that the
symmetrization operation can indeed be performed exactly.\ This is done by
introducing simple products of operators, corresponding to exchange cycles,
or more generally simple functions of operators (fractions) that correspond
to summations over the size of these cycles up to infinity. Physically, this
emphasizes the relation which exists between the properties of a system, its
pressure for instance, and the size of exchange cycles\cite{FL-1} that take
place in it.

Another feature of the method is that it naturally provides expressions for
the one and two body density operators, which is of course convenient if one
is interested in a detailed study of correlations.\ The formalism is useful
beyond the evaluation of virial corrections only; this is why we develop it
in what follows beyond what is strictly necessary to generalize the Beth
Uhlenbeck formula.\ A preliminary report on this work has already been given
in \cite{FL-1}; applications to denser systems will be the object of a
forthcoming article \cite{FL-2}.

\section{Ursell operators}

\ In the canonical ensemble, the partition function $Z_N$ of a system of
indistinguishable particles is given by a trace inside the space of
symmetrical (or antisymmetrical) states.\ Nevertheless, with the help of the
projectors $S$ and $A$ onto these (sub)spaces, the trace can be extended to
a larger space, which is associated with the same number of distinguishable
particles:
\begin{equation}
\label{20}Z_N=Tr\left\{ K_N\QATOP{S}{A}\right\}
\end{equation}
In this equation, $S$ applies for bosons while $A$ applies for fermions, and
the operators $K_N$ are defined by:
\begin{equation}
\label{21}
\begin{array}{ccl}
K_1 & = & \exp -\beta H_0(1) \\
K_2 & = & \exp -\beta \left[ H_0(1)+H_0(2)+V_{12}\right] \\
.... & ... & .... \\
K_N & = & \exp -\beta \left[ \sum_{i=1}^NH_0(i)+\sum_{i>j}V_{ij}\right]
\end{array}
\end{equation}
with usual notation: $H_0(1)$ is the one particle hamiltonian including
kinetic energy and an external potential (if there is any in the problem), $%
V_{ij}$ is the interaction potential between particles with labels $i$ and $%
j $. We now use cluster techniques to expand operators exactly in the same
way as one usually does for functions.\ The Ursell operators $U_l$ are
therefore defined according to:
\begin{equation}
\label{22}
\begin{array}{lll}
U_1 & = & K_1 \\
U_2(1,2) & = & K_2(1,2)-K_1(1)K_1(2) \\
U_3(1,2,3) & = & K_3(1,2,3)-K_2(1,2)K_1(3)-K_2(2,3)K_1(1) \\
&  & -K_2(3,1)K_1(2)+2K_1(1)K_2(2)K_1(3)
\end{array}
\end{equation}
etc. ($l=1,2,..N$). Conversely, in terms of the Ursell operators, the $N$
particle operator $K_N$ can be written in the form:
\begin{equation}
\label{23}K_N=\sum_{\left\{ m_l^{^{\prime }}\right\} }\sum_{\left\{
D^{^{\prime }}\right\} }\stackunder{m_1^{^{\prime }}\,factors}{\underbrace{%
U_1(.)U_1(.)...U_1(.)}}\times \stackunder{m_2^{^{\prime }}\,factors}{%
\underbrace{U_2(.,.)U_2(.,.)...U_2(.,.)}}\times U_3(.,.,.)\,....
\end{equation}
where the first summation is made on all possible ways to decompose the
number of particles as:
\begin{equation}
\label{23bis}N=\sum_llm_l^{^{\prime }}
\end{equation}
The second summation correspond to all non-equivalent\footnote{%
For instance $U_2(1,2)U_3(3,4,5)$ and $U_2(2,1)U_3(4,3,5)$ correspond to
equivalent distributions of 5 particles, since the very definition of the
Ursell operators implies that the order inside each $U_k$ is irrelevant.}
ways to distribute the $N$ particles into the variables of the Ursell
operators, symbolized by dots in (\ref{23}). It is convenient to simplify
the two summations into one:
\begin{equation}
\label{23ter}\sum_{\left\{ m_i^{^{\prime }}\right\} }\sum_{\left\{
D^{^{\prime }}\right\} }\Longrightarrow \sum_{\left\{ U\right\} }
\end{equation}
which is meant over all non-equivalent ways to distribute the particles into
various sequences of $U$'s. There is little difference between our
definitions and those of section 4.2 of \cite{Huang-chap14} or \cite{Pathria}%
: we use operators instead of symmetrized functions and, more importantly,
the action of these operators are defined, not only within the state space
that is appropriate for bosons or fermions, but also in the larger space
obtained by tensor product that occurs for distinguishable particles.\ Hence
the need for an explicit inclusion of $S$ or $A$ in (\ref{20}).

We now decompose these operators into a sum of permutations $P_\alpha $
that, in turn, we decompose into independent cycles $C$ of particles%
\footnote{%
We use the same notation as in ref.\cite{FL-1}: $C_k(i,j,k,..)$ denotes a
cycle where particle $i$ replaces particle $j$, particle $j$ replaces
particle $k$, etc. This should not be confused with the notation $%
P(i,j,k,...)$ for a general permutation (not necessarily a cycle) where
particle $i$ replaces particle 1, particle $j$ particle 2, particle $k$
particle 3, etc.\ Any $P$ can be decomposed into a product of $C$'s; for
instance, $P(2,1,3)=C_2(1,2)C_1(3)$.}:
\begin{equation}
\label{24}\QATOP{S}{A}=\frac 1{N!}\sum_{\left\{ m_i\right\} }\sum_{\left\{
D\right\} }\stackunder{m_1\,factors}{\underbrace{C_1(.)C_1(.)C_1(.)}}\times
\eta ^{m_2}\stackunder{m_2\,factors}{\underbrace{C_2(.,.)C_2(.,.)C_2(.,.)}}%
\times C_3(.,.,.)....
\end{equation}
where the first summation is similar to that of (\ref{23}), while the second
corresponds to all non-equivalent\footnote{%
Inside every cycle, a circular permutation of the variables has no effect
and therefore does not affect the permutation $P_\alpha $.} ways to
distribute numbers ranging from one to $N$ into the variables of the various
$C$'s.\ In this equation, the operator $S$ applies for bosons with $\eta =+1$%
, while $A$ with $\eta =-1$ applies for fermions. We also simplify the
notation into:
\begin{equation}
\label{24bis}\sum_{\left\{ m_i\right\} }\sum_{\left\{ D\right\}
}\Longrightarrow \sum_{\left\{ P_\alpha \right\} }
\end{equation}

We can now insert (\ref{23}) and (\ref{24}) into (\ref{20}) and obtain,
within a double summation, numbers that are traces calculated in the space
of distinguishable particles, i.e. in the ordinary tensor product of $N$
single particle state spaces.\ Inside most of the terms of the summation,
factorization into traces taken inside smaller subspaces occur.\ For
instance, if the term in question contains particle number $i$ contained at
the same time in a $U_1$ operator as well as in a $C_1$, the contribution of
that particle completely separates by introducing the simple number $%
Tr\left\{ U_1\right\} $. Or, if $n$ particles are all in the same $U_n$ but
all in separate $C_1$'s, this group of particles contributes by a factor $%
Tr_{1..n}\left\{ U_n\right\} ;$ if they are all in different $U_1$'s but
also contained in one single large cycle $C_n$, their contribution also
factorizes separately.\ More generally, in each term of the double
summation, particles group into clusters (U-C clusters), which associate
together all particles that are linked either\footnote{%
There are therefore two explicitly distinct origins to the clustering of
particles in this point of view: belonging to the same $U_{l^{^{\prime }}}$
(with $l^{^{\prime }}>1$) or to the same cycle $C_l$ (with $l>1$). This is
distinct from usual cluster theories where, either only interactions
introduce clustering (as in classical statistical mechanics), or the two
origins are not explicitely distinguished (as in usual quantum cluster
theory).} by cycles $C_l$ (with $l>1$) or by Ursell operators $%
U_{l^{^{\prime }}}$ (with $l^{^{\prime }}>1$).\ The general term is
therefore the product of the contributions of all the clusters that it
contains and one can write:
\begin{equation}
\label{25}Z_N=\frac 1{N!}\sum_{\left\{ P_\alpha \right\} }\sum_{\left\{
U\right\} }\prod_{clusters}\Gamma _{cluster}(i,j,k,..)
\end{equation}
where $(i,j,k,..)$ is the index number of particles contained inside the
cluster; the number of clusters into which each term of the double summation
is factorized depends, in general, on this particular term.

\section{Diagrams}

\ Clusters differing only by the numbering of particles that they contain
give the same contribution.\ It is therefore useful to reason in terms of
diagrams (U-C diagrams), which emphasize the way particles are connected
through exchange cycles and Ursell operators, rather than their numbering.
For instance, the first diagram will correspond to one particle in a $U_1$
and in a $C_1$ (whatever the numbering of the particle is), and contribute
the value $Tr\left\{ U_1\right\} $ as mentioned above; another diagram will
introduce the value $Tr_{1,2}\left\{ U_2\right\} $, etc.; more generally,
the value of any diagram containing $n_{diag.}$ particles can be written as
a trace over the variables of $n_{diag.}$ particles numbered arbitrarily:
\begin{equation}
\label{27b}
\begin{array}{cl}
\Gamma _{diag.}= & \eta ^{p_2+p_4+..}Tr_{1,2,..n_{diag.}}\left\{
U_1(1)\times \right. \\
& \,\,\,\,\,\,\,\,\,\,\,\left. \times
..U_1(r_1)U_2(r_1+1,r_1+2)...C_1(i)...C_1(j)C_2(l,s)...\right\}
\end{array}
\end{equation}
Here $p_k$ is the number of cycles\footnote{%
One obviously has: $p_1+2p_2+...=n_{diag.}$.} of length $k$; the factor $%
\eta ^{p_2+p_4+..}$ arises from the factors $\eta $'s in (\ref{24}) and
corresponds to the contribution of the parity of the permutations contained
in the diagram to the total permutation of the N particles.

Now, inside each term of the multiple summation, a given diagram $\Gamma
_{diag}$ may occur several times; we then note $m_{diag.}$ the number of
times it is repeated, and we get:
\begin{equation}
\label{26}Z_N=\frac 1{N!}\sum_{\left\{ P_\alpha \right\} }\sum_{\left\{
U\right\} }\prod_{diag.}\left[ \Gamma _{diag.}\right] ^{m_{diag.}}
\end{equation}
with the obvious relation:
\begin{equation}
\label{27}N=\sum_{diag.}m_{diag.}\times n_{diag.}
\end{equation}

Of course identical diagrams appear, not only in the same term of the double
summation, but also in many different terms.\ Therefore, if $\sum_{\left\{
m_{diag.}\right\} }$ symbolizes a summation over all possible ways to
decompose $N$ according to (\ref{27}), we can also write:
\begin{equation}
\label{28}Z_N=\frac 1{N!}\sum_{\left\{ m_{diag.}\right\} }c\left\{
m_{diag.}\right\} \times \prod_{diag.}\left[ \Gamma _{diag.}\right]
^{m_{diag.}}
\end{equation}
where $c\left\{ m_{diag.}\right\} $ is the numbers of terms in the double
summation of (\ref{26}) that correspond to this particular decomposition of $%
N$.

To evaluate this number, we have to specify more precisely how the U-C
diagrams are constructed graphically. In every cluster, we represent the
permutation cycles $C_k$ by horizontal lines containing $k$ boxes, or
segments, which are available to numbered particles.\ When the corresponding
particles are inside $U_1$'s, we do not add anything to the diagram; when
they are contained inside $U_2$'s, we join the corresponding segments by an
additional double line, a triple line for $U_3$'s, etc. For instance, if we
consider the pure exchange cluster:
\begin{equation}
\label{gamma1}\Gamma _{cluster}=Tr_{1,2,..,7}\left\{
U_1(1)U_1(2)..U_1(7)C_7(1,2,....7)\right\}
\end{equation}
the corresponding diagram will be that of figure 1-a.\ This kind of linear
diagram is the only possibility for an ideal gas; to generate a $\Gamma
_{cluster}$ appearing in (\ref{25}), it must receive a numbered particle in
the first segment of the line, in the second the particle which is replaced
by it under the effect of the permutation cycle, in the third the particle
which is replaced by that in the second segment, etc. Equation (\ref{gamma1}%
) gives what we will call the $``$explicit value'' of the contribution of
the diagram; using simple transformations (see ref.\cite{FL-1} or section
\ref{ideal}), one can obtain the simpler $``$reduced value'' $Tr_1\left\{
\left[ U_1(1)\right] ^7\right\} $.

\ If we now start from a cluster which contains one single $U_2$:
\begin{equation}
\label{gamma2}\Gamma _{cluster}=Tr_{1,2,.....9}\left\{
U_2(1,2)U_1(3)U_1(4)U_1(5)..U_1(9)C_3(1,3,4)C_6(2,5,....9)\right\}
\end{equation}
we obtain the diagram shown in figures 1-b; in the same way, the cluster:
\begin{equation}
\label{gamma3}\Gamma _{cluster}=Tr_{1,2,.....10}\left\{
U_2(1,6)U_1(2)U_1(3)..U_1(10)C_{10}(1,2,3,.,6,..10)\right\}
\end{equation}
leads to the diagram of fig. 1-c.\ These two kinds of diagrams turn out to
be the only ones that are necessary to generalize the Beth Uhlenbeck
formula; we give their reduced values in section \ref{second}. Fig. 1-d
shows an example of a diagram containing one single $U_3$, and arising from
the trace:
\begin{equation}
\label{gamma4}\Gamma =Tr_{1,2,.....,16}\left\{
U_3(1,2,3)U_1(4)U_1(5)..U_1(13)C_3(1,4,5)C_8(2,6..12)C_5(3,..16)\right\}
\end{equation}
and figure 1-e gives another similar example.\ Clearly, the process can be
generalized to associate a diagram to any cluster, however large and complex.

But it is not sufficient to construct diagrams, we must also choose explicit
rules\footnote{%
There is some flexibility in this choice, and here we attempt to take the
most convenient convention, but it is not necessarily the only possiblity.}
ensuring that every $\Gamma _{cluster}$ appearing in (\ref{25}), where the
particles are still numbered, will correspond to one single, well defined,
diagram; for instance, for the cluster written in (\ref{gamma2}), we must
decide whether the diagram will be that shown in figure 1-b or another where
the lowest cycle ($``$base cycle'') is that of length 6 instead of 3.\ We
will avoid this kind of ambiguity by choosing rules which fix, for every
cluster associated with a given diagram, where exactly each numbered
particle should fall into the diagram.\ These rules ensure that no double
counting of clusters may occur; they also determine how the geometrical
characteristics of the diagrams (the lengths of the successive cycles) can
be varied, which in turn determines the way in which the summations over
lengths will be made in a second step (next section). For our purpose in
this article, which is limited to a generalization of the Beth Uhlenbeck
formula and to the diagrams of figure 1-a,b and c, we do not need to study
the most general case (this discussion is given in Appendix I); it will be
sufficient to remember that the first particle in the lowest cycle ($``$base
cycle'') must be, among all particles contained in the $U_l$ of highest
order $l$, that which has the lowest index number for the configuration to
be correct.

The preceding rules also introduce the $``$counting factors'' corresponding
to the probability of obtaining a correct representation of a cluster by
throwing randomly numbered particles into a diagram; the counting factors
are used below to obtain the value of $c\left\{ m_{diag.}\right\} $. Suppose
for instance that we distribute $k$ numbered particles into a linear diagram
of the kind shown in fig.\ 1-a in all possible ways; it is clear that there
is a probability $1/k$ that the first particles will have the lowest index
number as required, which leads to the following counting factor $f_{diag.}$
for pure exchange cycles:
\begin{equation}
\label{1/k}f_{diag.}=\frac 1k
\end{equation}
(in other words, if one puts numbered particles into a cycle $C_k(.,.,.,..)$%
, one can obtain the same permutation $k$ different times). Similarly, it is
easy to see that the counting factors of the diagrams of figures 1-b and c
are $1/2$, which corresponds to the probability of having the particle
numbers contained inside the single $U_2$ in the correct order. The general
value of counting factors is given in Appendix I; in many practical
situations (when case (ii) of this Appendix does not occur) we can ignore
the $g$'s in formula (\ref{f'}) and use the simpler form:
\begin{equation}
\label{f}f_{diag.}=\left[ p_{l_M}\times l_M\right] ^{-1}
\end{equation}
where $l_M$ is the largest order of the Ursell operator contained in the
diagram (one for fig. 1-a, two for figs. b and c, three for figs. 1-d and e)
while $p_{l_M}$ is the number of these Ursell operators in this particular
diagram (7 for fig. 1-a, one for all the other cases in this figure).\ Two
other examples of diagrams for which this formula is valid are given in
figure 2; as an illustration, they are explicitly calculated in Appendix II.

We are now in position to calculate $c\left\{ m_{diag.}\right\} $. This
number can be obtained by distributing the $N$ particles inside the sites of
all the diagrams of a series defined by the $m$'s, which can be done in $N!$
different ways, and counting how many times the same term of the double
summation of (\ref{26}) is obtained.\ Since there are $m_{diag.}!$ ways to
interchange the order of all U-C clusters arising from the same diagram,
there is a first redundancy factor equal to $\prod_{diag.}\left(
m_{diag}!\right) $ that comes in.\ Moreover, there is also only a proportion
$\left( f_{diag.}\right) ^{m_{diag.}}$ of the obtained configurations that
is acceptable.\ Altogether, the net result is:
\begin{equation}
\label{29}c\left\{ m_{diag.}\right\} =N!\prod_{diag.}\frac
1{m_{diag.}!}\left[ f_{diag.}\right] ^{m_{diag.}}
\end{equation}

\section{Two summations\label{twosumma}}

We now take advantage of the fact that (\ref{29}) contains factorials and,
in a second step, that the value of $f_{diag.}$ depends on the topology of
the diagrams, but not of the size of the chains of $U_1$'s that it contains
(except for pure exchange sycles).\ This allows us to group together series
of terms in (\ref{28}).\ The first summation is done classically by going to
the grand canonical ensemble and defining the corresponding partition
function by:

\begin{equation}
\label{29b}Z_{g.c.}=\sum_N{\sl e}^{\beta \mu N}Z_N
\end{equation}
Then a useful simplification occurs because the sums over the $m_{diag.}$'s
are now independent; moreover the factors {\sl e}$^{\beta \mu N}$ can be
reconstructed by multiplying every number $\Gamma _{diag.}$ by {\sl e}$%
^{\beta \mu n_{diag.}}$, so that:
\begin{equation}
\label{30}Z_{g.c.}=\prod_{diag.}\exp \left[ \exp \left( \beta \mu
n_{diag.}\right) \times f_{diag.}\times \Gamma _{diag.}\right]
\end{equation}
We therefore obtain the grand potential (multiplied by $-\beta $) in the
form:
\begin{equation}
\label{31}\limfunc{Log}Z_{g.c.}=\sum_{diag.}{\sl e}^{\beta \mu
n_{diag.}}\times f_{diag.}\times \Gamma _{diag.}
\end{equation}
This is an exact formula, which gives the value of the pressure of the
system (multiplied by its volume and divided by the temperature).

The second summation which can now be done consists in grouping together the
contributions of all U-C diagrams that have the same $``$frame'' (or $``$%
skeleton); in other words we sum the diagrams which have the same topology
and differ only, inside the horizontal lines that represent the permutation
sycles, by the lengths of the intermediate chains of $U_1$'s that connect
together the particles contained in the $U_l$'s with $l>1$. Beside the fact
that this operation turns out to be mathematically simple - it merely leads
to the introduction of fractions of the operator $U_1$ as we will see in the
next section (see also the discussion of \S 2 of ref. \cite{FL-1}) -, it is
also indispensable from a physical point of view: we have to make a
summation over all lengths of intermediate horizontal chains of $U_1$'s in
order to take into account an arbitrary degree of degeneracy. We call $``$%
classes'' these groups of topologically equivalent U-C diagrams; classes may
also be represented by diagrams ($\Xi $-diagrams) which are, in a sense,
simpler than the original diagrams since any indication of the length of the
cycles has been removed.\ Examples are shown in figure 3 where dashed lines
emphasize that a summation over cycle length is implied.\ We call $\Xi
_{class}$ the contribution of a class:
\begin{equation}
\label{31bis}\Xi _{class}=\sum_{diag.\,\in \,class}{\sl e}^{\beta \mu
n_{diag.}}\times f_{diag.}\times \Gamma _{diag.}
\end{equation}
\ In terms of classes, (\ref{31}) becomes:
\begin{equation}
\label{31ter}\limfunc{Log}Z_{g.c.}=\sum_{classes}\Xi _{class}
\end{equation}

As (\ref{31}), this is an exact formula, containing extensive quantities in
both sides, and thus well adapted to approximations (as opposed to $Z_{g.c.}$%
). We will see below that the first term of the summation gives the grand
potential\footnote{%
More precisely, the logarithms of partition functions give the value of the
grand potential multiplied by $-\beta $.} of an ideal, (degenerate), gas,
which we shall note $\Xi _{ideal}$.\ The generalized Beth Uhlenbeck formula
is contained in the second and the third term in the summation, which we
shall note $\Xi _{direct}^1$ and $\Xi _{exch.}^1$; if, moreover, in each of
these two classes, one limits the summation of (\ref{31bis}) to its first
term (lowest order in $U_1$), one recovers the usual formula, valid only for
weakly degenerate gases.

\section{Dilute degenerate systems}

\subsection{Ideal gas\label{ideal}}

We first check that the first class of diagrams reconstructs the grand
potential (multiplied by $-\beta $) of the ideal gas. This class, symbolized
in figure 3-a, corresponds to the summation of the contribution of pure
exchange cycles containing only $U_1$'s, summed over any length $k$ ranging
from one to infinity. We know from (\ref{1/k}) the counting factor, so that
we just need to calculate the numerical contribution $\Gamma _k$ of every
cycle. The result, proved below, is simple:
\begin{equation}
\label{gamma}\Gamma _k=\eta ^{k+1}{\sl \,}\limfunc{Tr}\left\{ \left[
U_1\right] ^k\right\}
\end{equation}
where:
\begin{equation}
\label{eta}\eta =\QATOPD\{ \} {1\,\,\,\,\,{\sl for\,bosons}}{-1\,\,\,\,{\sl %
for\,fermions}}
\end{equation}
This is again a consequence of the fact that the trace over the $N$
particles contained in the diagram is taken in a space that is simply the
tensor product of $k$ single particle spaces of state. Because the numbering
of the particles does not affect the value of $\Gamma _k$ (it just changes
the names of dummy variables), we can for convenience renumber the relevant
particles from 1 to $k$.\ The effect of $C_k$ is then to move particle 1
into the place initially occupied by particle 2, particle 2 into the place
occupied by particle 3, and so on until one comes back to the place of
particle 1. Introducing a complete set of states $\left\{ \mid \varphi
_n>\right\} $ in the one particle space of states\footnote{%
If the particles have internal states, the index $n$ symbolizes at the same
time the orbital quantum numbers as well as those characterizing the
internal state.\ For instance, it the particles have spin $I$, a summation
written as $\sum_n$ contains in fact two summations, one over orbital
quantum numbers, and a second over $(2I+1)$ spin states.}, one can then
write:
\begin{equation}
\label{9ter}
\begin{array}{rcl}
\displaystyle\Gamma _k & \displaystyle= & \displaystyle\eta
^{k+1}\sum_{n_1,n_2,...n_k}<1:\varphi _{n_1}\mid U_1(1)\mid 1:\varphi
_{n_2}><2:\varphi _{n_2}\mid U_1(2)\mid 2:\varphi _{n_3}>\times \\  &  &
\displaystyle\times ....\times <k:\varphi _{n_k}\mid U_1(k)\mid k:\varphi
_{n_1}>\,=\eta ^{k+1}\,\limfunc{Tr}\left\{ \left[ U_1\right] ^k\right\}
\end{array}
\end{equation}
(the factor $\eta ^{k+1}$ is equal to the parity of the cycle which enters
in the definition of $A$ for fermions).\ We now have to make the summation:
\begin{equation}
\label{xi}\Xi _{ideal}=\sum_{k=1}^\infty \frac 1k{\sl e}^{\beta \mu k}\Gamma
_k=\sum_{k=1}^\infty \frac 1k{\sl e}^{\beta \mu k}\eta ^{k+1}{\sl \,}%
\limfunc{Tr}\left\{ \left[ U_1\right] ^k\right\}
\end{equation}
which contains a well known series:
\begin{equation}
\label{log}x+\eta \frac{x^2}2+\frac{x^3}3+\eta \frac{x^4}4+...=-\eta
\limfunc{Log}\left[ 1-\eta x\right]
\end{equation}
We therefore get for the grand potential (multiplied by $-\beta $) of the
ideal gas:
\begin{equation}
\label{32}\Xi _{ideal}=-\eta Tr\left\{ \limfunc{Log}\left[ 1-\eta e^{\beta
\mu }\,U_1\right] \right\}
\end{equation}
This is the classical result.\ For instance, we can assume that $H_0$, the
one-particle hamiltonian, is equal to $P^2/2m$ (kinetic energy of a particle
in a box); by replacing in (\ref{32}) the trace by a sum over $d^3k$, and $%
U_1$ by its diagonal element {\sl e}$^{-\beta \hbar ^2k^2/2m}$, one
immediately recovers usual formulas that are found in most textbooks on
statistical mechanics.\ Indeed, the method that we have used is more
indirect than the traditional method, but it gives a physical interpretation
to the term in $\left[ U_1\right] ^k$ that is obtained by expanding the
logarithmic function of (\ref{32}): it corresponds to the contribution of
all possible cyclic exchanges of $k$ particles in the system.

\subsection{Second virial correction\label{second}}

What happens now if we add the two following terms in (\ref{31ter}), which
contain one single $U_2$ and no Ursell operator of higher order ? Let us
start with the first class of diagrams, shown in fig. 3-b where the two
particles in the $U_2$ operator belong to two different exchange cycles, and
which we will call direct diagrams.\ In the first diagram of this class, the
two cycles are of length $k=1$ (identities) and only two particles,
unaffected by exchange, are involved; this simply introduces the
contribution:
\begin{equation}
\label{33}\Gamma _{11}=Tr_{1,2}\left\{ U_2(1,2)\right\}
\end{equation}
The next diagram in this class corresponds to three clustered particles, two
contained in the same $U_2$ and two in one permutation operator $C_2$.\ The
numerical value of this second diagram is:
\begin{equation}
\label{34}
\begin{array}{ccl}
\displaystyle\Gamma _{2,1} & \displaystyle= & \displaystyle\eta
Tr_{1,2,3}\left\{ U_2(1,2)U_1(3)C_2(1,3)C_1(2)\right\} \\  & = &
\displaystyle\,\eta Tr_{1,2,3}\left\{ U_2(1,2)U_1(3)P_{{\sl ex.}%
}(1,3)\right\}
\end{array}
\end{equation}
or:
\begin{equation}
\label{35}
\begin{array}{l}
\displaystyle\Gamma _{2,1}=\eta \sum_{n_1,n_2,n_3} \\ \,\,\,\,\,\,\,\,\,\,\,%
\,\,\displaystyle\,<1:\varphi _{n_1}\mid <2:\varphi _{n_2}\mid U_2(1,2)\mid
1:\varphi _{n_3}>\mid 2:\varphi _{n_2}><\varphi _{n_3}\mid U_1\mid \varphi
_{n_1}>
\end{array}
\end{equation}
which provides the following reduced value:
\begin{equation}
\label{36}\Gamma _{2,1}=\eta Tr_{1,2}\left\{ U_2(1,2)U_1(1)\right\}
\end{equation}
Similarly, one would calculate a contribution $\Gamma _{1,2}$ arising from
the exchange of particles 2 and 3, and obtained by replacing in (\ref{36}) $%
U_1(1)$ by $U_1(2)$. More generally, when $U_2$ clusters together $k_1$
particles, belonging to the same permutation cycle of length $k_1$, with $%
k_2 $ particles belonging to another cycle of length $k_2$, the calculation
of the effect of each of these cycles remains very similar to that of
section \ref{ideal}: now we have two particles that separately exchange with
others, but the algebra of operators remains the same for each of them. We
therefore get the reduced value:
\begin{equation}
\label{37}\Gamma _{k_1,k_2}=\eta ^{k_1-1}\eta ^{k_2-1}Tr_{1,2}\left\{
U_2(1,2)\left[ U_1(1)\right] ^{k_1-1}\left[ U_1(2)\right] ^{k_2-1}\right\}
\end{equation}
For this class of diagrams, according to (\ref{f}) the counting factor $f$
is simply $1/2$.\ The last step is to make a summation over all possible
values of $k_1$ and $k_2$ after inserting an exponential of $\beta $ times
the chemical potential multiplied by the number of particles contained in
the diagram:

\begin{equation}
\label{38a}\Xi _{direct}=\frac 12\sum_{k_1,k_2}{\sl e}^{\beta \mu \left(
k_1+k_2\right) }\Gamma _{k_1,k_2}\,\,\,\,
\end{equation}
\ This operation can be done by using the relation:
\begin{equation}
\label{homog}\sum_{k=1}^\infty \left( \eta x\right) ^{k-1}=\frac 1{1-\eta x}
\end{equation}
One therefore introduces fractions of the $U_1$'s operators, which results
in the expression:
\begin{equation}
\label{38b}\Xi _{direct}^1=\frac 12Tr_{1,2}\left\{ U_2(1,2)\frac{{\sl e}%
^{\beta \mu }}{1-\eta e^{\beta \mu }U_1(1)}\frac{{\sl e}^{\beta \mu }}{%
1-\eta e^{\beta \mu }U_1(2)}\right\}
\end{equation}

For the second class of diagrams, exchange diagrams shown in figure 3-c, the
two particles contained in $U_2$ are intermixed inside the same circular
permutation.\ The first exchange diagram corresponds to the two particles
contained in the same transposition:
\begin{equation}
\label{39}\Gamma _{1,1}^{{\sl ex.}}=Tr_{1,2}\left\{ U_2(1,2)C_2(1,2)\right\}
=Tr_{1,2}\left\{ U_2(1,2)P_{{\sl ex.}}\right\}
\end{equation}
The second will contain three particles:
\begin{equation}
\label{40}\Gamma _{1,2}^{{\sl ex.}}=Tr_{1,2,3}\left\{
U_2(1,2)U_1(3)C_3(1,2,3)\right\}
\end{equation}
which is equal to:
\begin{equation}
\label{41}
\begin{array}{ccl}
\displaystyle\Gamma _{1,2}^{{\sl ex.}} & \displaystyle= & \displaystyle%
\sum_{n_1,n_2,n_3}<1:\varphi _{n_1}\mid <2:\varphi _{n_2}\mid \\  &  &
\displaystyle U_2(1,2)\mid 1:\varphi _{n_2}>\mid 2:\varphi _{n_3}><\varphi
_{n_3}\mid U_1\mid 1:\varphi _{n_1}>
\end{array}
\end{equation}
Now, we can use the equality:
\begin{equation}
\label{42}
\begin{array}{l}
<1:\varphi _{n_1}\mid <2:\varphi _{n_2}\mid U_2(1,2)\mid 1:\varphi
_{n_2}>\mid 2:\varphi _{n_3}>= \\
\,\,\,\,\,\,\,\,\,\,\,\,\,\,\,\,\,\,\,\,\,\,\,\,\,\,\,\,\,\,\,\,\,\,\,\,\,\,%
\,\,\,\,\,\,\,<1:\varphi _{n_1}\mid <2:\varphi _{n_2}\mid U_2(1,2)P_{{\sl ex.%
}}\mid 1:\varphi _{n_3}>\mid 2:\varphi _{n_2}>
\end{array}
\end{equation}
which allows us to get the same summation over indices as in equation (\ref
{35}) and to obtain the reduced value:
\begin{equation}
\label{43}\Gamma _{1,2}^{{\sl ex.}}=Tr_{1,2}\left\{ U_2(1,2)P_{{\sl ex.}%
}U_1(1)\right\}
\end{equation}

Another, very similar, term occurs if the circular permutation $C_3(1,2,3)$
of (\ref{40}) is replaced by $C_3(1,3,2)$; the calculation can easily be
repeated and provides the result:
\begin{equation}
\label{44}\Gamma _{2,1}^{{\sl ex.}}=Tr_{1,2}\left\{ U_2(1,2)P_{{\sl ex.}%
}U_1(2)\right\}
\end{equation}

{}From the preceding equations it is not difficult to see that the generic
term of this second class of diagrams is obtained from (\ref{37}) by a
simple replacement of $U_2$ by the product $U_2P_{{\sl ex.}}$. Inserting a $%
P_{{\sl ex.}}$ into (\ref{38b}) therefore provides $\Xi _{exch{\sl .}}^1$
Finally, the value of the grand potential (multiplied by $-\beta $), to
first order in $U_2$, is given by:
\begin{equation}
\label{45}\limfunc{Log}Z_{g.c.}=\Xi _{ideal}+Tr_{1,2}\left\{ {\sl e}^{2\beta
\mu }U_2(1,2)\frac{\left[ 1+\eta P_{{\sl ex.}}\right] }2\frac 1{1-\eta
e^{\beta \mu }U_1(1)}\frac 1{1-\eta e^{\beta \mu }U_1(2)}\right\}
\end{equation}

This result is valid within an approximation which is basically a second
virial treatment of the interactions, while it contains all statistical
corrections.\ The formula remains therefore valid if the degree of
degeneracy of the gas is significant.\ Nevertheless, as pointed out for
instance in section 2.1 of ref. \cite{Huang-studies} and in ref. \cite
{Yang-Lee}, virial series (even summed to infinity) are no longer
appropriate beyond values where the density exceeds that of a phase
transition; this is because singularities in the thermodynamic quantities
occur at a transition (in the limit of infinite systems). Therefore, for
bosons, the validity of (\ref{45}) is limited to non condensed systems.\ The
discussion of what happens when a Bose Einstein condensation takes place
will be given in \cite{FL-2}.

Equation (\ref{45}) is a generalization of the Beth Uhlenbeck formula, and
reduces to it if the two denominators containing $U_1$'s are replaced by
one, an operation which is valid in the limit of low densities where ${\sl e}%
^{\beta \mu }$ is small. To see the equivalence between the low density
limit of relation (\ref{45}) with the usual value of the second virial
correction, we start from the definition of the $W_k$ functions given in
equation (14.35) of reference \cite{Huang-chap14}, which for $k=2$ becomes%
\footnote{%
We assume for simplicity that the particles have no internal state (or, if
they do, that they are all in the same internal state).}:
\begin{equation}
\label{a1}W_2({\bf r}_1,{\bf r}_2)=2\left( \lambda _T\right)
^6\sum_{n_s}\mid \Psi _{n_s}({\bf r}_1,{\bf r}_2)\mid ^2{\sl e}^{-\beta
E_{n_s}}
\end{equation}
where the functions $\Psi _{n_s}({\bf r}_1,{\bf r}_2)$ are a complete set of
stationary states for the system of identical particles (they are properly
symmetrized); the thermal wavelength is defined by:
\begin{equation}
\label{a1b}\lambda _T=\frac h{\sqrt{2\pi mk_BT}}
\end{equation}
Relation (\ref{a1}) can be transformed into:
\begin{equation}
\label{a2}
\begin{array}{rcl}
W_2({\bf r}_1,{\bf r}_2) & = & 2\left( \lambda _T\right) ^6<1:{\bf r}_1,2:%
{\bf r}_2\mid \sum_{n_s}\mid \Psi _{n_s}><\Psi _{n_s}\mid {\sl e}^{-\beta
H}\mid 1:{\bf r}_1,2:{\bf r}_2>
\end{array}
\end{equation}
Inside this equation appears a closure summation over the symmetrized states
of the system so that:
\begin{equation}
\label{a2bis}
\begin{array}{rcl}
W_2({\bf r}_1,{\bf r}_2) & = & 2\left( \lambda _T\right) ^6<1:{\bf r}_1,2:%
{\bf r}_2\mid S\,{\sl e}^{-\beta H}\mid 1:{\bf r}_1,2:{\bf r}_2>
\end{array}
\end{equation}
(for fermions, $S$ is replaced by $A$). From this function ref. \cite
{Huang-chap14} defines the Ursell function $U_2^H({\bf r}_1,{\bf r}_2)$ by:
\begin{equation}
\label{a3}U_2^H({\bf r}_1,{\bf r}_2)=W_2({\bf r}_1,{\bf r}_2)-W_1({\bf r}%
_1)W_1({\bf r}_2)=\Delta W_2({\bf r}_1,{\bf r}_2)+U_2^{ideal\,gas}({\bf r}_1,%
{\bf r}_2)
\end{equation}
where $\Delta W_2({\bf r}_1,{\bf r}_2)$ is the difference between the values
of $W_2({\bf r}_1,{\bf r}_2)$ with and without interaction potential, and $%
U_2^{ideal\,gas}({\bf r}_1,{\bf r}_2)$ the value of the Ursell function for
a system of two free particles. Equation (14.49)\footnote{%
In the second edition, this equation is numbered (10.49).} of \cite
{Huang-chap14} shows that the second virial coefficient in the expansion of
the grand potential is half of the integral of $U_2^H({\bf r}_1,{\bf r}_2)$
over the variables ${\bf r}_1$ and ${\bf r}_2$, multiplied by $\left(
\lambda _T\right) ^{-3}$ and the inverse of the volume.\ But the
contribution of $U_2^{ideal\,gas}({\bf r}_1,{\bf r}_2)$ is automatically
contained in $\limfunc{Log}Z_{g.c.}$, so that we can concentrate on $\Delta
W_2({\bf r}_1,{\bf r}_2)$ only.\ Because the summation over ${\bf r}_1$ and $%
{\bf r}_2$ can be written as a trace in a space which is the tensor product
of two one particle state spaces, and because:
\begin{equation}
\label{a4}S,A=\frac 12\left[ 1\pm P_{{\sl ex.}}\right]
\end{equation}
we get the result:
\begin{equation}
\label{a5}\int {\sl d}^3r_1\,{\sl d}^3r_2\,\Delta W_2^V({\bf r}_1,{\bf r}%
_2)=2\left( \lambda _T\right) ^6Tr_{1,2}\left\{ U_2(1,2)\frac{\left[ 1+\eta
P_{{\sl ex.}}\right] }2\right\}
\end{equation}
where $U_2(1,2)$ is the difference between the exponentials of the
interacting particle hamiltonian minus that of free particles, which is
precisely our definition (\ref{22}).\ We therefore recover our result (\ref
{45}), provided the denominators $1-\eta {\sl e}^{\beta \mu }U_{1\text{ }}$%
inside the trace are replaced by one.

\subsection{Discussion; reduced density operators}

The only difference between equation (\ref{45}) and the Beth Uhlenbeck
formula arises from the presence of the two fractions inside the trace.
Since:

\begin{equation}
\label{U1}\frac 1{1-\eta {\sl e}^{\beta \mu }U_1}=1+\eta \frac{{\sl e}%
^{\beta \mu }U_1}{1+\eta {\sl e}^{\beta \mu }U_1}
\end{equation}
they are actually nothing but operatorial forms of the usual Fermi or Bose
factors $\left( 1+\eta f\right) $ that appear, for instance, in the
collision term of the Uehling Uhlenbeck or Landau kinetic equation (with the
usual notation $f$ for the distribution function) .\ In the present case,
nevertheless, because $U_2$ and $U_1$ do not commute in general, the
operatorial character of the fractions is relevant.\ For free (but mutually
interacting) particles in a box, the eigenvectors of $U_1$ are plane waves%
\footnote{%
If the particles are subject to the effect of an external potential (atoms
in a trap for instance), the eigenvectors of $U_1$ are not simple plane
waves, but the essence of our analysis remains valid.}, while those of $U_2$
are different since they involve correlations between the particles.

If the trace in the right hand side of (\ref{45}) is calculated in the basis
of plane waves, the correction is expressed as an integral containing the
diagonal elements of $U_2$ between such plane waves.\ The simplest situation
occurs when all these diagonal elements have the same sign.\ Then, for
bosons, because the eigenvalues of the fractions are larger than $1$, the
effect of degeneracy is always to enhance the effects of interactions; this
is physically satisfying since the Bose Einstein statistics tends to favor
situations where particles are close.\ In particular, if the potential is
attractive and if there are two body bound states (molecules), their weight
will be increased with respect to what is would be in the usual Beth
Uhlenbeck formula.\ If, on the other hand, the diagonal elements of $U_2$
between plane waves change sign when the relative momentum of the two
particles changes, which may happen if the potential has attractive as well
as repulsive parts, more complicated cancellation effects may take place in
both the usual Beth Uhlenbeck formula and its generalization, so that no
general prediction on the sign of the effect of degeneracy is possible
(except for the contribution of bound states which remains enhanced as
above).

For fermions, the eigenvalues of the fractions are between $0$ and $1$ so
that the effect of statistics are just the opposite of what they are for
bosons: they tend to reduce the effects of the interactions, except if
mutual cancellation effects take place.\ Moreover, if the system is strongly
degenerate, the effect of the product of the two fractions is to cancel the
contribution of all matrix elements corresponding to particles inside the
Fermi sphere, leaving only interactions between particles near the surface
(or outside) of the Fermi sphere.\ This applies to bound molecules, which
introduce a contribution containing the scalar product of the bound state
wave function by all plane waves outside of the Fermi sphere, exactly as in
the Cooper problem.

The applicability of (\ref{45}) is not limited to the equation of state of
the quantum system, it can also be applied to the evaluation of partial
density operators, for instance the one particle density operator. If we set%
\footnote{%
If one wishes to introduce hermitian operators only, one can symmetrize (\ref
{u1}) by putting square roots of $U_1$'s on both sides of $\overline{U\text{.%
}}$}:
\begin{equation}
\label{u1}U_l(1,2,..l)=\left[ U_1(1)\times U_1(2)\times ...\times
U_1(l)\right] \overline{U}(1,2,..l)
\end{equation}
we can choose $U_1$ and the $\overline{U}_l$'s, instead of the $U_l$'s, to
characterize the correlations between the particles; by definition $%
\overline{U}_1$ is equal to one.\ Assume now that we keep all $\overline{U}%
_l $'s constant and vary $U_1$ according to:
\begin{equation}
\label{u2}dU_1=d\varepsilon \mid \varphi ><\varphi \mid U_1
\end{equation}
What is then the variation of $Z$? Inserting this relation into (\ref{u1}),
then into (\ref{23}) and finally into (\ref{20}), we get:

\begin{equation}
\label{14}dZ_N=d\epsilon \,\,Tr_{1,2,3,...N}\left\{ \QATOP{S}{A}%
K_N\sum_{i=1}^{i=N}\mid i:\varphi \rangle \langle i:\varphi \mid \right\}
\end{equation}
This expression is nothing but the product of $Z_N$ by the average value of
the one particle operator $\sum_{i=1}^{i=N}\mid i:\varphi \rangle \langle
i:\varphi \mid $. Therefore, in terms of the one particle density operator%
\footnote{%
Here we take the convention where the operator $\rho _I$ is normed to the
number $N$ of particles (instead of one).} $\rho _I$, we obtain:
\begin{equation}
\label{14b}\frac d{d\epsilon }\limfunc{Log}Z_N=NTr\left\{ \mid \varphi
\rangle \langle \varphi \mid \rho _I\right\} =N\langle \varphi \mid \rho
_I\mid \varphi \rangle
\end{equation}
(the derivative with respect to $\epsilon $ is meant at the value $\epsilon
=0$). We can therefore obtain reduced density matrices from this kind of
operatorial variations.\ If the grand canonical ensemble is used, the
many-particle density operator is defined by:
\begin{equation}
\label{15}\rho _{gc}=\left[ Z_{gc}\right] ^{-1}\sum_N{\sl e}^{\beta \mu
N}\left\{ \QATOP{S}{A}K_N\QATOP{S}{A}\right\}
\end{equation}
and we find the following relation involving the one particle density
operator $\rho _I^{gc}$:
\begin{equation}
\label{16}\frac d{d\epsilon }\limfunc{Log}Z_{gc}=Tr\left\{ \rho _I^{gc}\mid
\varphi \rangle \langle \varphi \mid \right\} =\langle \varphi \mid \rho
_I^{gc}\mid \varphi \rangle
\end{equation}
This formula shows that $\rho _I^{gc}$ is the $``$operatorial derivative''
of $Z_{gc}$ (we have only considered diagonal variations of $U_1$ but, since
the basis of states $\mid \varphi \rangle $ is arbitrary, a generalization
to off-diagonal variations is easy).

For the ideal gas, all $\overline{U}_l$ are zero if $l\geq 2.$ We can then
simply use (\ref{32}) and take a derivative with respect to $\varepsilon $
by choosing in succession for $\mid \varphi \rangle $ all eigenvectors of $%
\rho _F$ .\ In this way we obtain:
\begin{equation}
\label{17}\rho _I^{gc}=\frac{{\sl e}^{\beta \mu }U_1}{1-\eta {\sl e}^{\beta
\mu }U_1}
\end{equation}
which is a well known result. When interactions are added, one should
replace (\ref{32}) by (\ref{45}) with the substitution (\ref{u1}) for $U_2$%
.\ One then gets a modified expression of $\rho _I$ where contributions
arise from the derivatives of the denominators in (\ref{45}) as well as that
of the numerator (through the variation of $U_2$), that we do not write here
for brevity.\ This shows how the method is adaptable to the evaluation of
microscopic quantities; if one is interested in microscopic correlations,
two body density operators can be obtained by a similar method. This is a
specificity of our technique, as compared for instance to that based on the
pseudopotential; the latter is built in order to reconstruct the eigenvalues
of the hamiltonian, and hence the partition function, but not the wave
functions, in particular certainly not at short relative distances.

A final remark is related to the convergence of the power series that we
have summed into fractions of the $U_1$ operators (the remark applies for
fermions only). As noted by Kahn and Uhlenbeck \cite{Kahn-Uhlenbeck}, when
the chemical potential of a system of fermions becomes positive, the virial
series diverge, and the equation of state is obtained by a continuation of a
an analytic function.\ Here we observe the same phenomenon: the series in $%
\left[ {\sl e}^{\beta \mu }U_1\right] ^k$ that we have summed over the size $%
k$ of the exchange cycles becomes divergent when $\mu >0$; nevertheless the
sum remains a regular function and, for fermions, (\ref{45}) has no
singularity.

\section{Conclusion}

The basic idea of the method of Beth and Uhlenbeck, or more generally of
Mayer cluster expansion methods \cite{Mayer}, is to reason in terms of
functions which directly give local approximations of the thermal
equilibrium.\ Here we start from Ursell operators which give rise to various
contributions in the form of U-C diagrams.\ This leads to equation (\ref
{31ter}) which provides an exact expression of the grand potential of the
system, and expresses it as sum of various terms arising from $\Xi $%
-diagrams, already containing a sum over all possible sizes of intermediate
exchange cycles.\ Each term is obtained as an integral (a trace) over a
finite number of variables. The general expression is valid for dilute or
dense systems as well, such as liquids or even solids, but for gases it
becomes simpler because it can be truncated more abruptly. Indeed, the
generalization of the Beth Uhlenbeck formula is obtained by limiting the
summation to the first three terms only (the first corresponding to the
ideal gas).\

In its spirit, the method that we have used is close to the $``$binary
collision approximation of Lee and Yang \cite{Lee-and-Yang}\cite
{Pathria-bincoll}''; they also introduce the Ursell functions of an
auxiliary system obeying Boltzmann statistics, and they establish relations
between the Ursell functions of the two systems (see also ref. \cite{Mohling}
for a discussion of this type of method).\ But these relations acquire a
rapidly increasing complexity with the order of the Ursell functions, and
since they do not use Ursell operators, it becomes necessary at some point
to make approximations (integrations over the inverse temperature $\beta $%
).\ In our method, we introduce operators which generalize Ursell functions,
and we make no attempt to evaluate the Ursell functions of the system of
identical particles.\ Rather, we calculate the complete contribution of each
initial (unsymmetrized) Ursell operator to the symmetrized partition
function; this explains why the final results are markedly different.

Another point of comparison is the more recent calculation by Nozi\`eres and
Schmitt Rink \cite{Schmitt}, who give a calculation of the thermodynamic
potential for a system of fermions at low densities.\ In order to simplify
the summation of the diagrams in their calculation, they assume that the
matrix elements of the interaction potential are separable into a product of
functions; moreover, for brevity, they explicitly include only one phase
shift (s wave approximation) so that their result is less general than (\ref
{45}).\ Of course this does not mean that their method can not be
generalized to fully recover (\ref{45}), but we have not examined the
question. In the same vein, a more detailed comparison between (\ref{45})
with the results of Galitskii \cite{Galitskii} for fermions and Belyaev \cite
{Belyaev} for bosons, valid a zero temperature, would be useful.

\begin{center}
ACKNOWLEDGMENTS
\end{center}

Part of this work was done during a pleasant and interesting visit in Leiden
at the Kamerling Onnes Laboratory; the author is very grateful to his host
Prof.\ Giorgio Frossati, as well to his co-visitor Prof.\ Sandro Stringari,
for discussions and intellectual stimulation. Several useful and friendly
discussions with Philippe Nozi\`eres are also acknowledged.\ The text of
this article owes a lot to Prof.\ W.\ Mullin and to Peter Gr\"uter, who both
suggested corrections and improvements.

\

\newpage

\begin{center}
APPENDIX\ I : GENERAL RULES\ FOR\ CONSTRUCTING DIAGRAMS
\end{center}

Our convention for building the diagram associated to any particular term in
the double summation (\ref{25}) is the following:

(i) we start from the $U_l$, or the $U_l$'s, that are of highest order $%
l=l_M $ in this particular term, and identify the particle it (or they)
contain(s) that has the lowest index number $n_{\min }$; this particle is
considered as the $``$base particle'', and belongs to the $``$base cycle''
from which all the rest of diagram will be drawn. By convention, the base
particle is shown first (left position) and the base cycle is that at the
lowest position in the figure; for instance, in figures 1-b and 1-d, this
base is a three particle cycle.

(ii) to continue in the construction of the diagram, we add a second
generation of cycles.\ We first use the Ursell operator $U_{l_M}$ that
contains the base particle and add the other permutation cycles which
include the other particles in the same $U_{l_M}$; this is done in the order
of increasing values for the index number of the particles contained in this
$U_{l_M}$, so that the order of the new cycles is clearly defined.\ This
also defines, for each of them, a $``$secondary base particle'' that is put
first in the diagram\footnote{%
If the connection is multiple, that is if the additional cycle contains
several particles of the same base Ursell operator $U_{l_m}$, the secondary
base particle is that of lowest index number; all the other particles are
then automatically located in the diagram by their order in the exchange
cycle, and do not play a special role at this stage.}.

(iii) we continue the addition of this second generation of branches in the
diagram by moving along the base cycle and skipping all numbered particles
that are in $U_1$'s, until we reach one which belongs to an $U_l$ with $%
l\geq 2$; we then add additional cycles containing the other particles
inside this $U_l$. We use the same rule an in (ii) and we define $``$%
secondary bases'' for the new cycles, so that their representation is also
uniquely fixed. Going along all the base cycle in this way completes the
first generation of additional cycles.

(iv)\ Then we build in the same way the second generation, by starting in
succession from all of cycles of the first generation in the order in which
they were added, etc.\ until, eventually, the complete diagram is obtained.

With these conventions, all cycles are individually identified, so that it
makes sense to vary their lengths independently to generate all terms of (%
\ref{25}); this will be useful below for the calculations which lead from
the U-C to the $\Xi $-diagrams. Figure 2 shows examples of diagrams;
Appendix II gives more details on the conventions used in their
representation (in particular, it is convenient to assume that, inside the
trace, the series of Ursell operators are put before the series of cycles).

Now suppose that we reverse the question: starting from a given diagram, how
do we identify its occurrence in the double sum?\ what is its weight in the
summation that gives $\limfunc{Log}Z$? Assume that we throw randomly $%
n_{diag.}$ numbered particles into all available locations. With the precise
rules that we have chosen above, it is clear that double counting problems
are avoided, but also that not all of these random configurations obtained
are allowed.\ What is the proportion $f_{diag.}$ of the configurations that
are compatible with our conventions? Let us note $p_{l_M}$ the total number
of operators $U_{l_M}$ (those of largest order $l_M$) appearing in this
particular diagram. The reasoning is as follows:

(i) first there is a probability $\left( p_{l_M}\times l_M\right) ^{-1}$
that the right base particle will be obtained

(ii) second, if $l_M\geq 3$, in the construction of successive generations
of cycles, either more than one cycle is added from the connections of the
same $U_l$, or there is a multiple connection towards the same cycle (or
both). Figure 1-e gives one example of such a multiple connection.\ In all
these cases, additional factors $\left( g\right) ^{-1}$ are introduced which
account for the correct ordering of numbering of secondary base particles.

Thus we obtain:
\begin{equation}
\label{f'}f_{diag.}=\left[ p_{l_M}\times l_M\times g\right] ^{-1}
\end{equation}
(this formula is also valid if $l_M=1$, in which case $p_{l_M}$ is nothing
but the size $k$ of the linear exchange cycle). For the generalization of
the Beth Uhlenbeck formula, case (ii) never happens and the factors $g$'s do
not play any role; they would nevertheless enter the calculations in
general, for instance in the discussion of Bose Einstein condensation of
groups of three particles in \cite{FL-2}.

\begin{center}
APPENDIX II : TWO\ EXAMPLES
\end{center}

\ The diagram shown in figure 2-a corresponds by definition to the following
trace:
\begin{equation}
\label{x1}Tr_{1,2,3,4,5}\left\{
U_2(1,2)U_2(3,5)U_1(4)C_1(1)C_3(2,3,4)C_1(5)\right\}
\end{equation}
Our convention is that the Ursell operators are always put before the
cycles; the notation $C_3(2,3,4)$ refers to a cycle where the particle
numbered $2$ replaces that numbered $3$, that numbered $3$ replaces that
numbered $4$, and that numbered $4$ that numbered $2$ (the $C_1$ 's do not
produce any change in the positions of the particles). Equation (\ref{x1})
gives what we call the $``$explicit value'' of this particular diagram, but
it can also be simplified into a $``$reduced value''.\ This can be done by
inserting into (\ref{x1}) closure relationships and using summations to
introduce products of operators whenever possible.\ In this case this leads
to the expression:
\begin{equation}
\label{x2}Tr_{1,2,5}\left\{ U_2(1,2)U_2(2,5)U_1(2)\right\}
\end{equation}
In the explicit value, any numbered particle appears once and only once in
every Ursell operator and every cycle; in the reduced form this is not
necessarily the case.

In a similar way the diagram shown in figure 2-b is defined by the explicit
expression:
\begin{equation}
\label{x3}Tr_{1,2,3,4,5,6}\left\{
U_2(1,2)U_1(3)U_2(4,5)U_1(6)C_1(1)C_3(2,3,4)\eta C_2(5,6)\right\}
\end{equation}
while its reduced value is:
\begin{equation}
\label{x4}\eta Tr_{1,2,5}\left\{ U_2(1,2)U_1(2)U_2(2,5)U_1(5)\right\}
\end{equation}

\smallskip

\begin{center}
FIGURE\ CAPTIONS
\end{center}

Fig.\ 1: Examples of U-C\ diagrams.\ For an ideal gas, only linear U-C
diagrams containing chains of one particle Ursell operators $U_1$ occur, as
shown in (a); a summation of the contribution of these diagrams over the
length of the chain gives the grand potential (multiplied by $-\beta $).\
The generalization of the Beth Uhlenbeck formula arises for the diagrams
shown in (b) and (c) containing one single two body Ursell operator $U_2$;
for each of them, a summation over the lengths of the $U_1$ chains is also
necessary.\ Figure (d) shows an example of a diagram containing a three body
$U_3$ operator and three $U_1$ chains, with a counting factor $g=2$. Figure
(e) shows an example of multiple connection through an $U_3$ operator, which
also introduces a factor $g=2$ in the weight of the diagram. For more
details on the definition and counting of the diagrams, see Appendix I.%
\smallskip

Fig.\ 2: Other examples of more complex U-C diagrams.\ As in figure 1,
horizontal lines correspond to exchange cycles containing $U_1$ operators;
in addition, vertical double lines symbolize $U_2$ operators (triple lines
would be used for $U_3$, etc.). Formula (\ref{f}) gives the weight of these
diagrams; a more explicit calculation of these two terms is given in
Appendix II.\smallskip\

Fig.\ 3: $\Xi $ diagrams introduced by the summation of the diagrams of
figure 1 over the lengths of the $U_1$ chains, according to formula (\ref
{31bis}). The dashed lines, which symbolize these summations, can be
replaced by intermediate operators given by fractions $1/\left[ 1-\eta
zU_1\right] $, where $z=\limfunc{exp}\beta \mu $; every operator $U_l$, with
$l\geq 2$, remains explicit and, moreover, introduces a factor $z^l$;
finally, the weights $f_{diag.}$ must also be inserted in the value of the $%
\Xi $ diagram.

\begin{center}
BIBLIOGRAPHY
\end{center}

\ \

\end{document}